----------------------------------------------------------------------------
\documentstyle[12pt,setspace,fullpage]{article}
\doublespace
\newcommand{\be}{\begin{equation}}
\newcommand{\ee}{\end{equation}}
\newcommand{\beq}{\begin{eqnarray}}
\newcommand{\eeq}{\end{eqnarray}}
\advance\voffset by -1truecm

\begin{document}

\begin{center}
{\Large{The Mass Quantum and Black Hole Entropy}} \\
\vskip 2em
{\large{\sf B. RAM}} \\ [2mm]
Physics Department, New Mexico State University, \\
Las Cruces, NM 88003, USA.
\end{center}
\vskip 4em

\doublespacing 

\begin{abstract}
We give a method in which a quantum of mass equal to twice the Planck
mass arises naturally. Then using Bose-Einstein statistics we derive
an expression for the black hole entropy which physically tends to the
Bekenstein-Hawking formula.
\end{abstract}

\vskip 4em
\hrule width 3cm
\medskip

\newpage

With the introduction by Bekenstein \cite{jdb73} of the intriguing
concept of black hole entropy and that it is proportional to the black
hole area $A$, followed by the work of Bardeen, Carter and Hawking
\cite{jmb73} and of Hawking \cite{swh75} (see also \cite{rmw75,lp75}),
it now appears that $\kappa/2\pi$ is the physical temperature $T$ of a
black hole, $\kappa$ being its surface gravity. In order to make this
firm, it therefore becomes imperative that a calculation
of $S$ from first principles in quantum statistics be made and that it
give the Bekenstein-Hawking relation $S=A/4$. 

In a recent publication Ashtekar {\it et al} \cite{aa98}, using a
quantum theory of geometry, obtain, for a large non-rotating
black hole, an expression for $S$ which reduces to $A/4$ with a
specific choice of the Immirzi parameter \cite{gi97} which enters the 
calculation. The {\it horizon} plays an important role in their
method. (For other attempts at calculating the black hole
entropy, see refs. \cite{aa98,cr96} and references cited therein.)

In the present paper we give an alternative derivation of $S$ for a
Schwarzschild black hole based on quantum statistics and obtain an
expression which tends to the Bekenstein-Hawking relation. Throughout
this paper we deal with only the Schwarzschild case, and use units in
which $G=c=\hbar= k =1$.

Another promising concept which was introduced by Bekenstein
\cite{jdb74} is that squared irreducible mass $M^2_{ir}$ of a black
hole may be quantized in the form
\[
M^2 = gn \ , \hspace{1cm} n = 1,2, \cdots 
\]
$(M_{ir}=M$ for a Schwarzschild black hole) with $g$ a constant. The
alternative form $A = \alpha n$ is used in
refs.\cite{vfm86,jdb95,sh98} as $A = 16 \pi M^2$. What struck us most
about this particular quantization rule of Bekenstein was that it
strongly implied the {\it presence} of a {\it simple harmonic
oscillator}, {\it somewhere} in the quantum dynamics of a black
hole. Below we show that the harmonic oscillator is {\it truly}
present. 

Let us start with the classical timelike geodesic equation
\cite{rmw84}
\be
{1\over 2} \ \dot r^2 + {L^2\over 2r^2} - {m\over r} - {mL^2 \over r^3} 
= {1\over 2} (E^2 - 1) 
\ee
of a test particle, where an overdot represents differentiation with
respect to $\tau$ the proper time, $m$ is the mass that enters the
Schwarzschild metric \cite{rmw84}
\be
ds^2 = - \left(1 - {2m\over r}\right) dt^2 + \left(1 - {2m\over
r}\right)^{-1} dr^2 + r^2 d\Omega^2 ,
\ee
$E = \left(1-{2m\over r}\right) \dot t$, a conserved quantity, is
interpreted as the total energy per unit rest mass of the test
particle following the geodesic \cite{rmw84}, and $L = r^2 \dot\phi$, 
also a conserved quantity, is the angular momentum per unit rest mass
of the test particle. Equation (1) shows that it is the same
\cite{rmw84} as that of a unit mass particle in non-relativistic
mechanics \cite{hg50} with kinetic energy
\[
K \ E = {1\over 2} \ \dot r^2 + {L^2\over 2r^2} ,
\]
effective potential energy
\[
V = - {m\over r} - {mL^2\over r^3} ,
\]
and effective total energy
\[
\bar E = {1\over 2} (E^2 - 1) .
\]
Hence, using the usual Schr\"odinger prescription \cite{em70} in (1),
we obtain the corresponding stationary state quantum equation
\be
\left(- {1\over 2} \nabla^2 - {m\over r} - {mL^2\over r^3}\right) \psi 
= {1\over 2} (E^2-1) \psi .
\ee
Eq. (3), in $r, \theta, \phi$ coordinates, is
\be
\left[- {1\over 2r^2} \ {\partial \over \partial r} \left(r^2 {\partial
\over \partial r}\right) + {L^2\over 2r^2} - {m\over r} - {mL^2\over
r^3}\right] \psi = {1\over 2} (E^2 - 1) \psi .
\ee
With
\[
\psi = R(r) Y^m_\ell (\theta, \phi)
\]
the radial part of Eq. (4) is given by
\be
\left(- {1\over 2r^2} \ {d\over dr} \left(r^2 {d\over dr}\right) + {\ell 
(\ell + 1)\over 2r^2} - {m\over r} - {m\ell (\ell + 1) \over
r^3}\right) R = {1\over 2} (E^2-1) R .
\ee
Note that the quantum equation (5) is characterized by $m, E$ and
$\ell$. 

Now let us {\it bring in the horizon}. This is done by letting the
variable $r$ be the coordinate of the horizon and the operator
$m(r)=r/2$. $m$ varies as $r$ varies. Substituting $r/2$ for $m$ 
in the last term on the LHS of Eq. (5) makes the term
equal to ${\displaystyle{-\ell (\ell + 1) \over 2r^2}}$ which cancels
the second term on the LHS, ${\displaystyle{+\ell (\ell + 1) \over
2r^2}}$. Physically this means that the horizon has ``swallowed'' all
information about the angular momentum of the test particle. 
In the third term $-m/r$, however, instead of replacing $m$ by $r/2$,
we replace it by its eigenvalue which is denoted by the same symbol
$m$. (That $m$ is indeed the eigenvalue becomes evident as one proceeds.)
And we get the equation
\be
\left(- {1\over 2r^2} \ {d\over dr} \left(r^2 {d\over dr}\right) -
{m\over r}\right) R = {1\over 2} (E^2 - 1) R .
\ee
To ensure that {\it no trace} of the test particle is left
and that the variable $r$ {\it is} the horizon coordinate, 
we put $E
= 0$ and obtain
\be
\left(- {1\over 2r^2} \ {d\over dr} \left(r^2 {d\over dr}\right) -
{m\over r}\right) R = - {1\over 2} R 
\ee  
as the quantum equation that is characterized by only $m$,
the Schwarzschild mass. (It is easy to see now that if we had replaced
the operator $m$ in the term $-m/r$ by $r/2$ rather than its eigenvalue,
there would be no $m$ left to quantize and
we would not be able to proceed any further.)
With $m = \mu/4$ and $U= r R(r)$, Eq. (7)
reduces to the simple form
\be
\left(- {1\over 2} \ {d^2\over dr^2} - {\mu/4 \over r}\right) 
U = - {1\over 2} U . 
\ee 
Eq. (8) is the quantum eigenvalue equation for the mass $\mu$. That
$\mu$ has the eigenvalues
\be
\mu = 2 (n+1) \omega
\ee
with $\omega = 2$, $n = 0, 1, 2, \cdots$ is seen via the straight
forward method of Mavromatis \cite{ham97}:

The radial Schr\"odinger equation for the N-dimensional oscillator
problem can be written as \cite{ac90} 
\beq
&& \left[- {1\over 2} \left({d^2\over dr^2} - {\left(\ell + {N\over 2} -
{3\over 2}\right) \left(\ell + {N\over 2} - {1\over 2}\right) \over
r^2}\right) + {1\over 2} \omega^2 r^2\right] U^{(N)}_{n\ell} (r)
\nonumber \\  
&& \hspace{5cm} = \left(2n + \ell + {N\over 2}\right) \omega U^{(N)}_{n\ell} (r) ,
\eeq
where $n= 0,1, \cdots$, and the orbital quantum number $\ell$ is a
positive interger or zero. In $N'$ dimensions the Coulomb counterpart
of Eq. (10) is
\be
\left[- {1\over 2} \left({d^2\over ds^2} - {\left(\ell' + {N'\over 2} -
{3\over 2}\right) \left(\ell' + {N'\over 2} - {1\over 2}\right) \over
s^2}\right) - {\alpha \over s}\right] \psi (s) = - B \psi (s) ,
\ee 
$N'$ being the dimension of the Coulomb space, $\ell'$ the
corresponding orbital angular momentum quantum number, and $B$ the
Coulomb binding energy. Under the transformations
\be
s = \rho^2 , \ \ \ \ \ {\rm and} \ \ \ \ \ \psi = \rho^{1/2} \phi ,
\ee
Eq. (11) becomes
\be
\left[- {1\over 2} \left({d^2\over d\rho^2} - {\left(2\ell' + {2N'-2\over 2} -
{3\over 2}\right) \left(2\ell' + {2N'-2\over 2} - {1\over 2}\right) \over
\rho^2}\right) + 4B\rho^2 \right] \phi (\rho)  = 4 \alpha \phi (\rho) .
\ee
Comparison of Eqs. (10) and (13) shows that there is a satisfactory
mapping when
\beq
\ell = 2\ell' \ \ &,& \ \ N = 2N' - 2 , \nonumber \\
{1\over 2} \omega^2 = 4B \ \ &,& \ \ (2n + \ell + {N\over 2}) \omega =
4\alpha .
\eeq

Now {\it notice} that our Eq. (8) is Eq. (11) with
\be
\ell' = 0 , \ \ N'=3 , \ \ B = {1\over 2} , \ \ \alpha = \mu/4 ;
\ee
and hence it represents a {\it four-dimensional harmonic oscillator}
with $\omega = 2$, and $\mu$ given by Eq. (9). In modern language
Eq. (9) or
$$
\mu_n = 2 (n+1) \omega \ , \ \ \ \ n = 0, 1, 2, \cdots 
\eqno(9')
$$
says that the n-th mass $(\mu)$ state is occupied by $n$ {\it pairs}
of mass quanta, each quantum being of $\omega = 2$ or mass equal to
{\it twice} the Planck mass. This means that there are no mass states
with odd number of quanta.
 
Now for calculating the entropy of a black hole of mass $M$, let $2N$
mass quanta be enclosed in a volume $V$ and their total mass
be $M = 2 N\omega$. Let us rewrite Eq. ($9^\prime$) as
\be
\epsilon_n = {\mu_n \over 2} = (n +1) \omega , \ \ \ n = 0, 1, 2,
\cdots 
\ee
so that (i) the nth $\epsilon$-state contains $n$ quanta, (ii) the
thermodynamic quantity $E$ is given by
\be
E = {M\over 2} = N \omega ,
\ee
and (iii) the thermodynamic probability of the macroscopically defined state
$(N, E)$ can simply be calculated by literally applying Bose's
method as given in his original paper \cite{bose24}.

Let $p_0$ be the number of vacant $\epsilon$-cells, $p_1$ the number
of those $\epsilon$-cells which contain one quantum, $p_2$ the number
of $\epsilon$-cells containing two quanta, and so on. Then the
probability of the state defined by the $p_r$ is obviously
\be
W = {P! \over p_0! p_1! \cdots}
\ee
where
\be
  P = \sum_r p_r
\ee
is the total number of $\epsilon$-cells\footnote{The total number of
$\mu$-cells is also $P$. The only difference between an
$\epsilon$-cell and a $\mu$-cell is that if an $\epsilon$-cell
contains $r$ quanta, then the corresponding $\mu$-cell contains $2r$
quanta.} over which
\be
N = \sum_r r p_r 
\ee
quanta are distributed. Since $p_r$ are large numbers, we have
\be
\ell n W = P \ell n P - \sum_r p_r \ell n p_r .
\ee
Then it is straightforward \cite{bose24} to maximize (21) satisfying 
the auxiliary conditions (17) and (20), and one obtains
\beq
p_r &=& P (1 - e^{-\omega/\beta}) e^{-r \omega/\beta} \ , \\ [2mm]
N &=& P (e^{\omega/\beta} - 1)^{-1} \ , 
\eeq
and
\be
S = {E\over \beta} - P \ell n (1 - e^{-\omega/\beta}) .
\ee
>From the condition ${\partial S \over \partial E} = {1\over T}$, one
obtains $\beta = T$; and (24) becomes
\be
S = {E\over T} - P \ell n (1 - e^{-\omega/T}) .
\ee
For a Schwarzschild black hole for which $T = {\kappa \over 2\pi} =
{1\over 8\pi M}$ and $A = 16\pi M^2$, (25) becomes
\be
S = {A\over 4} - {M\over 4} (e^{A/M} - 1) \ell n (1 - e^{-A/M})
\ee
which for large $M$ physically\footnote{Mathematically Eq. (26)
reduces to ${\displaystyle{S = {A\over 4} + {M\over 4}}}$.} tends to
\be
S = {A\over 4} .
\ee
Thus the picture of a black hole that emerges from above is:

A black hole is a Bose-Einstein ensemble of quanta of mass equal to
twice the Planck mass, confined in a volume of two-sphere of radius
twice the black hole mass. 

The author thanks A. Ram and A. Dabholkar for useful conversations, N.
Ram and R.S. Bhalerao for critical reading of the manuscript and
making concrete suggestions, and the Tata Institute for a pleasant
stay. The author also thanks the referee for asking two questions
which led to further clarification of the method.

\end{document}